# Per-Instance Algorithm Selection for Recommender Systems via Instance Clustering

Andrew Collins

   ADAPT Centre, School of Computer Science and Statistics, Trinity College Dublin, Dublin, Ireland, ancollin@tcd.ie

Laura Tierney

   School of Mathematics, Trinity College Dublin, Dublin, Ireland, tiernela@tcd.ie

Joeran Beel

   ADAPT Centre, School of Computer Science and Statistics, Trinity College Dublin, Dublin, Ireland & The University of Siegen, Department of Electrical Engineering and Computer Science, Germany. j@beel.org

**ABSTRACT**

Recommendation algorithms perform differently if the users, recommendation contexts, applications, and user interfaces vary even slightly. It is similarly observed in other fields, such as combinatorial problem solving, that algorithms perform differently for each instance presented. In those fields, meta-learning is successfully used to predict an optimal algorithm for each instance, to improve overall system performance. Per-instance algorithm selection has thus far been unsuccessful for recommender systems. In this paper we propose a per-instance meta-learner that clusters data instances and predicts the best algorithm for unseen instances according to cluster membership. We test our approach using 10 collaborative- and 4 content-based filtering algorithms, for varying clustering parameters, and find a significant improvement over the best performing base algorithm at $\alpha = 0.053$ (MAE: 0.7107 vs LightGBM 0.7214; t-test). We also explore the performances of our base algorithms on a ratings dataset and empirically show that the error of a perfect algorithm selector monotonically decreases for larger pools of algorithm. To the best of our knowledge, this is the first effective meta-learning technique for per-instance algorithm selection in recommender systems.

## 1  Introduction

Recommendation algorithms perform differently for varying scenarios [10], contexts [4], datasets [11], applications [6], and user characteristics [5, 17]. Even slight changes in these can impact recommender system performance. Practitioners tend to address this difficulty by evaluating several algorithms, to identify an overall-best single algorithm to use for a system. A single overall-best algorithm can provide an adequate performance when compared to other single algorithms, but it will not be optimal for all instances.

Meta-learning is an approach used to automatically predict an optimal algorithm for each data instance. Meta-learners learn the relationship between characteristics of data and the performance of algorithms on that data. Per-instance meta-learning, wherein the best algorithm is predicted for each data instance, is not common in recommender systems, and what little research exists shows mixed results. It is successfully applied in other domains however, such as: combinatorial search problem solving [24], where each instance is a descriptor for a single problem to be solved; computer game solving [8] where each instance describes the state of a game; pattern recognition [1] where each instance might be an image that must be classified.

In this paper we propose a per-instance meta-learner for recommender systems. To the best of our knowledge this is the first example of an effective per-instance meta-learner for a recommendation dataset.

We cluster data instances in a MovieLens dataset and identify the best algorithm for each cluster from a pool of 10 collaborative- and 4 content-based filtering algorithms. For unseen instances we predict a best algorithm according to cluster membership. We furthermore examine algorithm performances per-cluster and suggest ways in which such an analysis may be useful for algorithm selection researchers generally. We empirically show how a hypothetically perfect meta-learner performs according to the number of base algorithms that are available to it. Lastly, we discuss why we think our proposed approach works, when compared to other algorithm selectors and when compared to our own previous approaches.

## 2 Related Work

For meta-learners, clustering can be used for all parts of the architecture. For example, it is used to derive meta-features for per-dataset meta-learning models. Instances within a dataset are clustered, and useful descriptions of these clusters are identified, for example, the number of clusters generated, the proportion of instances in clusters, or the distance between centroids [27][33]. These meta-features are then related to algorithm performances and can be used for algorithm selection. Clustering is further used for base algorithms that are then recommended by a meta-learner, specifically for clustering tasks [18][35][31]. The meta-models in these cases are not cluster-based algorithms, but rather regressors, neural networks, and decision trees. Lastly clustering is also used as a meta-model. Lee and Giraud-Carrier [26] cluster algorithms according to the similarity of their behaviour on a given instance. Instances are described by a set of 22 statistical and landmarking meta-features. For new instances, a cluster of similarly-behaving algorithms is identified, and a random algorithm from this cluster is selected. They compare this approach to a more common meta-learning method of classifying a single predicted-best algorithm and show a 12% improvement in classification accuracy.

In all related work mentioned so far, meta-learning and clustering are used in predicting a single algorithm for future datasets (e.g., Lee and Giraud-Carrier [26] use 85 datasets, and therefore have 85 instances for cross-validation of their meta-learner).

Per-instance meta-learning is not common for recommender systems. See Collins et al. [11] and Collins and Beel [9] for examples of per-instance meta-learning using linear regression and random forests for error prediction, and Ekstrand and Riedl [16] as an example using logistic regression for per-subset algorithm classification. Siamese Neural Networks are also used for per-instance algorithm selection in recommender systems [7, 36].

Edenhofer et al. [15] cluster data instances using k-means for a recommendation task. Four base algorithms are used for each data instance, and their performance according to their recall in a top-N set is noted. Each cluster is assigned a best-algorithm according to majority voting. They cluster new instances and identify a predicted best-algorithm according to the training data. Clustering is the least effective of all meta-learning approaches (decision tree, gradient boosting, stacking using decision trees), and is 22.8% worse than the best base algorithm (content-based filtering using TF-IDF).

Nechaev et al. [28] use the HDBSCAN algorithm to cluster user features and item features in 5 recommendation datasets. They identify the cost that is incurred when a meta-learner selects a non-optimal algorithm. Their meta-learner therefore chooses between the predicted-best algorithm per-user, and the overall-best algorithm, according to a utility function that factors in errors caused by their meta-learner. Their clustering approach improves RMSE compared to the overall-best algorithm for 2 out of 5 datasets, with a greatest improvement of 3% (Modcloth dataset, RMSE; 0.95 for a KNN base algorithm vs 0.914 for user clustering). Their worst result is for MovieLens100K, for which clustering performs 0.11% worse than the overall best algorithm.

We have previously examined meta-learners using performance estimation and error prediction as the target of our meta-model [9, 11] to mixed results. For example, we predicted the best algorithm per-instance from a pool of collaborative filtering algorithms using two MovieLens datasets (100K and 1M). Our error estimators (using random forest regressors) were unable to discriminate algorithm performance adequately (RMSE, 100K: 0.973; 1M: 0.908) and performed 2-3% worse than overall-best single algorithm SVD++ (RMSE, 100K: 0.942; 1M: 0.887). This was due in-part to the similarity in performance and prediction for the top algorithms in the pool; it is relatively easy to predict the worst algorithms in a pool, but if some subset of algorithms perform similarly across all instances, choosing the best one approaches a random selection and performance converges to the average performance of all algorithms.

Cunha et al. [12–14] conduct a literature review and several experiments for per-dataset algorithm selection using meta-learning, for recommender systems. Kotthoff also conducts a literature review of per-instance meta-learning for combinatorial search problems [24] and Smith-Miles reviews of per-instance meta-learning in numerous fields [34]. Otherwise, clustering techniques have long been used in algorithm selection literature, unrelated to meta-learning; see Rendell et al. [32], Houstis et al. [20], Aha et al. [2], Kuncheva [25] for early examples.

To the best of our knowledge, clustering has not been used effectively for per-instance algorithm selection in recommender systems.

## 3 Methodology

We want to predict the optimal algorithm, from a pool of algorithms, for each instance in a recommendation dataset. Our algorithm selector can choose from 14 recommendation algorithms in all. We use these algorithms to make rating predictions for MovieLens100K instances, noting that this dataset is commonly used for recommender system evaluations [3] but that no meta-learning approaches have been positively evaluated for any MovieLens dataset, as far as we are aware (see Collins et al. [11], Ekstrand and Riedl [16], Nachaev et al. [28]).

Ten of our 14 algorithms are collaborative filtering variants, from the Surprise library [21] (listed in Figure 1). These algorithms use the user id, item id, and rating from each row in the dataset. We furthermore use linear regression (LR) using SciKit [30] and LightGBM [22] (LGBM). LR and LightGBM use standard dataset features: user id, gender, age, occupation, user zip code, movie genre and release date. We augment the original data by replacing the zip code with the median income for that zip code using U.S. Census data. We convert the movie release date from DD/MM/YY format to year only. We one-hot encode categorical variables, and standardize continuous variables using z-score.

We use two more variations of LR and LightGBM with extended features (LR_EF, LGBM_EF), using statistical meta-features for each row: mean/standard deviation/minimum/maximum/median, and the number of ratings, for each user and each item. The fully augmented dataset comprises 58 features.

### 3.1 Algorithm Performance Analysis

We aim to identify subgroups in the data within which data instances have a similar best algorithm, according to their performance in rating prediction. To do this we cluster data instances using k-means clustering, using Euclidean distance to measure instance similarity.

For the fully augmented variant of the dataset with all 58 features, Euclidean distance will result in circular clusters that may contain dissimilar instances. To prevent this, we use Principal Component Analysis (PCA) to remove collinearity between variables and reduce dimensions to 3.

We select the number of clusters to use for k-means by minimizing Within-Cluster Sum of Squared Error (WSS). We calculate WSS for a range of k and select the 'elbow point'; for our data, k=10.

We use Mean Absolute Error (MAE) to assess algorithm performance. The most effective algorithm for an instance is that with the smallest MAE.

### 3.2 Meta-Learning Oracles

We do not know how an algorithm selector will perform according to the number of base algorithms, and strength of base algorithm, that are available to it. To assess whether using any of our pool of base algorithms with a meta-learner could be effective, we calculate a hypothetically optimal MAE value for combinations of algorithms. For each algorithm in the combination, we predict a rating for each instance. For a given instance, we log the minimum error from the combination of algorithms. We calculate the mean of these values across all instances as our overall MAE for this combination of algorithms.

## 3.3 Algorithm Selection Implementation

For each cluster generated by k-means, we calculate the MAE for all 14 algorithms of the clustered instances. The algorithm with the overall smallest MAE is chosen to be the most effective algorithm for that cluster and this algorithm is noted for future use when selecting a rating-prediction algorithm for new instances. We retrain any promising algorithms based on our performance analysis (section 3.1).

Our algorithm selector finds the nearest cluster for unseen instances using k-Nearest Neighbor (KNN). We then use the overall-best algorithm for this cluster on the new instance. For each new instance, the predicted class by KNN is the cluster to which the instance is predicted to belong. The previously identified best-performing algorithm for each cluster is then used to predict the rating for this instance. This is done for each instance in the test set. We calculate the total MAE for each algorithm selector.

We compare the total MAE to 1) the MAE of the mean of the predictions from all base algorithms that can be selected, and to 2) the MAE of the overall best-performing algorithm.

Using 5-fold cross validation, we include 100K instances from the dataset.

## 4 Results

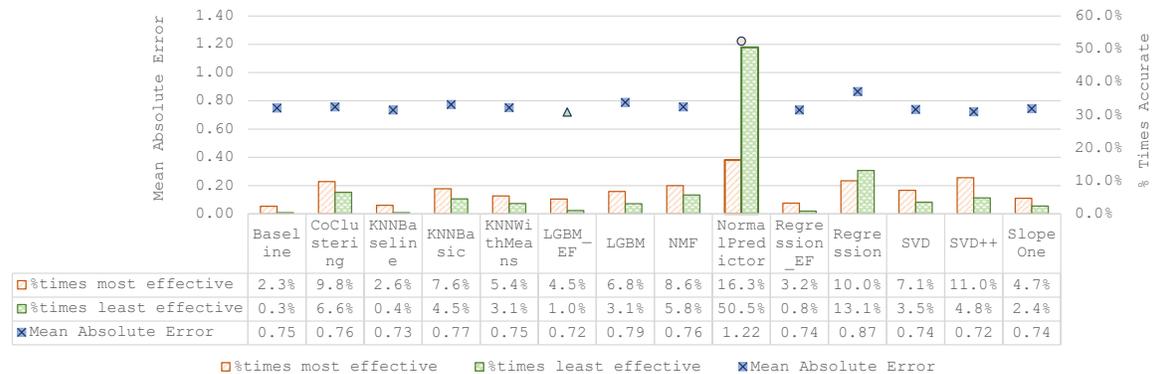

| | Baseline | CoClustering | KNNBaseline | KNNBasic | KNNWithMeans | LGBM_EF | LGBM | NMF | NormalPredictor | Regression_EF | Regression | SVD | SVD++ | SlopeOne |
|---|---|---|---|---|---|---|---|---|---|---|---|---|---|---|
| %times most effective | 2.3% | 9.8% | 2.6% | 7.6% | 5.4% | 4.5% | 6.8% | 8.6% | 16.3% | 3.2% | 10.0% | 7.1% | 11.0% | 4.7% |
| %times least effective | 0.3% | 6.6% | 0.4% | 4.5% | 3.1% | 1.0% | 3.1% | 5.8% | 50.5% | 0.8% | 13.1% | 3.5% | 4.8% | 2.4% |
| Mean Absolute Error | 0.75 | 0.76 | 0.73 | 0.77 | 0.75 | 0.72 | 0.79 | 0.76 | 1.22 | 0.74 | 0.87 | 0.74 | 0.72 | 0.74 |

**Figure 1: Mean Absolute Errors for each base algorithm, and the percentage of instances for which each algorithm is most and least effective. Normal Predictor is most frequently both the most and least effective algorithm, but also has the worst MAE. LGBM_EF is most effective according to MAE.**

### 4.1 Base Algorithm Effectiveness

Figure 1 shows the MAE of all base algorithms. LGBM_EF is the overall best-performing algorithm with a MAE of 0.72. From these results, if we had to choose one algorithm to use for this data or for a recommender system with similar data, we would choose LGBM_EF. The top 5 overall best-performing algorithms have MAE values within a small range (within 0.017). LGBM_EF is an improvement of just 0.2% on the next best-performing algorithm (SVD++).

LGBM_EF, which is the overall best-performing algorithm (MAE 0.72), is most effective for just 4.3% of all instances. Despite having the best MAE it is not the best available algorithm for most of the instances (95.7%). This tells us that LGBM_EF performs consistently well, but that just using this overall best-performing algorithm for this dataset is not optimal.

Normal Predictor is least effective overall (MAE: 1.22) and is 41% worse than the second least effective algorithm, Linear Regression (0.74). We expect that Normal Predictor would be comparatively poor as it is a basic algorithm that predicts a random rating according to an assumed-normal distribution of ratings in the training dataset. Normal Predictor is most frequently the most effective algorithm (16.3%, 16,300 instances), despite having the worst MAE (Figure 1). Normal Predictor is also most often the least effective algorithm (50.5%, ~50,500 instances). Given that Normal Predictor is the algorithm with the worst MAE value (1.22), many instances where Normal Predictor is the least effective algorithm is to be expected.

## 4.2 Meta-Learner Oracle

Table 1 lists the MAE for combinations of algorithm in a meta-learning oracle. MAE for a perfect meta-learner that combines the two most frequently effective algorithms (Normal, SVD++; Figure 1) could decrease MAE 21.09% over the single best base algorithm (0.563, compared to LGBM_EF: 0.721). Effectiveness could improve with each successive additional algorithm up to an MAE of 0.353 for 14 algorithms. From 5 algorithms onwards this improvement in MAE plateaus, with a decrease of 13.1% in MAE between using 5 algorithms and 14 algorithms. Perfectly combining all 14 algorithms MAE would result in an MAE 0.353, an improvement of 51.0% over LGBM_EF.

This result is evidence of high *performance complementarity* [23] among our choice of base algorithms. That is, our base algorithms are each effective for distinct subsets of instances; they have complementary strengths in the problem space.

MAE for a perfect meta-learner decreases monotonically for larger pools of base algorithm and converges to some minimum optimal value. This corresponds with analyses of classifier ensembles in pattern recognition, where generalization error for ensembles decrease similarly [19][29]. However, weighted ensembles tend to put more emphasis on more effective algorithms and therefore are not overly affected by poor base algorithms. In the case of single-best algorithm selectors, it might be expected that out-of-sample error increases for larger pools of algorithms due to overfitting of the meta-model, and due to the penalty that a non-optimal algorithm selection carries. As such, Table 1 suggests that a trade-off exists between the potential gains that a perfect algorithm selector presents, and the errors that an imperfect algorithm selector will make for larger numbers of base algorithms. The asymptote in Table 1 may give an indication of the maximum number of base algorithms to use for a generalizable single algorithm selector.

**Table 1. MAE for combinations of base algorithms, from least to most effective. MAEs are calculated using a rating prediction 'oracle' that selects the best rating prediction per row. Combinations are selected according to % most effective (Figure 1)**

| Algorithm Combinations | MAE |
|---|---|
| LGBM_EF | 0.721 |
| NP, SVDpp | 0.563 |
| NP, SVDpp, LR | 0.485 |
| NP, SVDpp, CoC, LR | 0.432 |
| NP, NMF, SVDpp, CoC, LR | 0.406 |
| NP, NMF, KNNBasic, SVDpp, CoC, LR | 0.390 |
| NP, NMF, SVD, KNNBasic, SVDpp, CoC, LR | 0.379 |
| NP, NMF, SVD, KNNBasic, SVDpp, CoC, LGBM_EF, LR | 0.370 |
| NP, NMF, SVD, KNNBasic, SVDpp, CoC, LGBM_EF, LGBM, LR | 0.364 |
| NP, NMF, KNNWithMeans, SVD, KNNBasic, SVDpp, CoC, LGBM_EF, LGBM | 0.360 |
| NP, NMF, KNNWithMeans, SVD, KNNBasic, SVDpp, SlopeOne, CoC, LGBM_EF, LGBM,LR | 0.356 |
| NP, NMF, KNNWithMeans, SVD, KNNBasic, SVDpp, SlopeOne, CoC, LGBM_EF, LGBM, Reg_EF,LR | 0.355 |
| NP, NMF, KNNWithMeans, SVD, KNNBasic, KNNBaseline, SVDpp, SlopeOne, CoC, LGBM_EF, LGBM, Reg_EF,LR | 0.354 |
| BaselineOnly, NP, NMF, KNNWithMeans, SVD, KNNBasic, KNNBaseline, SVDpp, SlopeOne, CoC, LGBM_EF, LGBM, Reg_EF,LR | 0.353 |

## 4.3 K-means Cluster Results

We analyze clusters from k-means to assess whether algorithms perform distinctly for different groups in the dataset. We chose the number of clusters in the dataset to be 10, based on Within-Cluster Sum of Squared Error and an 'elbow' plot. We assign labels 1-10 to each cluster.

For 6 clusters out of 10, SVD++ has the smallest MAE (Table 2). The median MAE difference between SVD++ for these 6 clusters and the algorithm with the second smallest MAE is 0.009. For the remaining 4 clusters out of 10, LGBM_EF has the smallest MAE. The median MAE difference between LGBM_EF for these 4 clusters and the algorithm with the second smallest MAE is 0.021; when LGBM_EF is the algorithm with the smallest MAE, it has a much larger median difference between it and the second best algorithm, compared to SVD++ (50.7% larger). When SVD++ has the smallest MAE, LGBM_EF is next most effective for 3 clusters, KNNBaseline for two clusters and SVD for the remaining one cluster. When LGBM_EF has the lowest MAE, LR_EF had the second-smallest MAE for three out of four clusters and KNNWithMeans had the second-smallest MAE for the remaining cluster. In other words, when SVD++ has the smallest MAE, LGBM_EF is the next best algorithm 50% of the time. However, when LGBM_EF is the algorithm with the smallest MAE, SVD++ is never the next best. We also see that when SVD++ is not the algorithm with the smallest MAE, LGBM_EF is, and vice versa.

**Table 2. The most effective (bolded) and second most effective (italicized) algorithm in each cluster according to MAE**

| Cluster: | 0 | 1 | 2 | 3 | 4 | 5 | 6 | 7 | 8 | 9 |
|---|---|---|---|---|---|---|---|---|---|---|
| **Baseline Only** | 0.622 | 0.818 | 0.803 | 0.651 | 0.750 | 0.894 | 0.705 | 0.808 | 0.630 | 0.888 |
| **CoClustering** | 0.626 | 0.835 | 0.854 | 0.632 | 0.769 | 0.861 | 0.720 | 0.818 | 0.622 | 0.964 |
| **KNN Baseline** | 0.608 | *0.801* | 0.810 | 0.637 | 0.740 | 0.852 | 0.688 | 0.795 | *0.614* | 0.894 |
| **KNN Basic** | 0.622 | 0.832 | 0.853 | 0.715 | 0.753 | 0.919 | 0.691 | 0.841 | 0.623 | 1.482 |
| **KNNWith Means** | 0.622 | 0.821 | 0.830 | *0.627* | 0.758 | 0.852 | 0.725 | 0.812 | 0.633 | 0.945 |
| **LGBM_EF** | **0.589** | 0.805 | *0.789* | **0.622** | *0.737* | **0.793** | 0.698 | *0.788* | 0.620 | **0.639** |
| **LGBM** | 0.639 | 0.857 | 0.832 | 0.753 | 0.778 | 0.967 | 0.741 | 0.832 | 0.669 | 0.832 |
| **NMF** | 0.633 | 0.825 | 0.840 | 0.656 | 0.762 | 0.878 | 0.709 | 0.806 | 0.640 | 0.951 |
| **Normal Predictor** | 1.139 | 1.250 | 1.179 | 1.173 | 1.110 | 1.531 | 1.161 | 1.196 | 1.215 | 1.968 |
| **LR_EF** | *0.597* | 0.822 | 0.808 | 0.629 | 0.744 | *0.826* | 0.708 | 0.803 | 0.618 | *0.791* |
| **LR** | 0.712 | 0.939 | 0.869 | 0.854 | 0.812 | 1.076 | 0.832 | 0.887 | 0.754 | 1.348 |
| **SVD** | 0.620 | 0.805 | 0.803 | 0.644 | 0.746 | 0.867 | *0.684* | 0.794 | 0.628 | 0.876 |
| **SVD++** | 0.612 | **0.791** | **0.787** | 0.628 | **0.734** | 0.832 | **0.669** | **0.777** | **0.610** | 0.869 |
| **Slope One** | 0.619 | 0.814 | 0.825 | 0.636 | 0.754 | 0.856 | 0.692 | 0.808 | 0.620 | 0.971 |

## 4.4 Selecting Algorithms with a Meta-Learner

Following our analysis of algorithm effectiveness per-cluster (Table 2) we manually inspect our data and see that SVD++ and LGBM_EF have complimentary performance for this dataset. They would be sensible base algorithms to use in a minimal meta-learner. The hypothetically perfect MAE for this combination of algorithms results in a MAE of 0.6063.

We use KNN to predict the cluster classification for each new instance and our meta-learner selects the best algorithm from SVD++ and LGBM_EF for this cluster. This meta-learner produces an MAE of 0.7142 (Table 3), which is an improvement of 0.0072 (1%) over the most effective algorithm for the test set (LGBM_EF: 0.7214). This improvement is not statistically significant (t-test; $\alpha = 0.05$, p=0.1942). This meta-learner is further 1.03% more effective than a weighted ensemble of LGBM_EF and SVD++ (MAE: 0.7216). Instances for which LGBM_EF was used had an overall MAE of 0.6725, and likewise, SVD++ had an overall MAE of 0.7218.

We test other clustering parameters to evaluate meta-learning with more granularity within clusters. The most effective algorithms per-cluster varies when 25 clusters are used; LGBM_EF had the smallest MAE for 12 clusters (48%); SVD++ for 9 (36%); KNNBasic for 2 (8%); Co-Clustering for 1 (4%); and Slope One for 1 (4%). We expect that a more diverse range of best-performing algorithms within the cluster will provide a better overall MAE. We therefore further evaluate a meta-learner using these 5 base algorithms with 25 clusters. This gives an MAE of 0.7128 (Table 3), which is an improvement of 0.0086 (1.19%) over the single-best algorithm on the test set (LGBM_EF; MAE: 0.7214). The difference in MAE between this meta-learner and LGBM_EF is not statistically significant (t-test; $\alpha = 0.05$, p=0.1216). This meta-learner gives an improvement of 0.0322 (4.32%) over a weighted ensemble of the 5 base algorithms.

Using 5 algorithms and 25 clusters improves MAE just 0.0014 compared to 2 algorithms and 10 clusters, so it is not clear that using all 5 algorithms is advantageous. We therefore lastly investigate MAE using the two most frequently effective algorithms for 25 clusters: SVD++ and LGBM_EF. This meta-learner produces an MAE of 0.7107 (Table 3), which is an improvement of 0.0107 (1.48%) over the single-best algorithm on the test set (t-test; p=0.053). This meta-learner also improves MAE over a weighted ensemble of the base algorithms by 1.51%.

It is noteworthy that using 25 clusters with 2 algorithms is more effective than using 5 algorithms. This could be caused by underfitting, affecting the generalizability of the model.

**Table 3. Meta-Learner results, compared to the single-best base algorithm, using 2 or 5 base algorithms, and 10 or 25 clusters**

| Algorithm | MAE | SD | MAE average ensemble | Improvement vs best algo. | Improvement % | p-value |
| --- | --- | --- | --- | --- | --- | --- |
| LGBM_EF | 0.7214 | 0.5528 | | - | - | - |
| Meta-Learner 2/10 | 0.7142 | 0.5561 | 0.7216 | 0.0072 | 1.01% | 0.1942 |
| Meta-Learner 5/25 | 0.7128 | 0.5579 | 0.7450 | 0.0086 | 1.21% | 0.1216 |
| Meta-Learner 2/25 | 0.7107 | 0.5539 | 0.7216 | 0.0107 | 1.51% | 0.0532 |

## 5 Conclusion

To the best of our knowledge this is the first example of a successful per-instance meta-learner for a recommendation dataset. In the approach we propose here, we can see distinct performances for algorithms across clusters (Table 2) that, overall, have the same MAE (e.g., SVD++ and LGBM_EF (Figure 1)) and can discard algorithms that do not have distinct performances even though they are often effective (e.g., Normal predictor). Removing non-complementary algorithms from your pool likely reduces the difficulty in predicting a best algorithm from a pool of strong algorithms that we noted from previous work.

In future work we hope automatically designate complementary algorithms and discard non-complementary algorithms based on their performances across clusters, and extend the analysis across more datasets. Even though k-means seems sufficient, other clustering approaches should also be examined, such as spectral clustering, as well as non-Euclidean distance measures.

### ACKNOWLEDGMENTS
This publication has emanated from research conducted with the financial support of Science Foundation Ireland (SFI) under Grant Number 13/RC/2106 and funding from the European Union and Enterprise Ireland under Grant Number CF 2017 0303-1.

### REFERENCES
[1] Aguiar, G.J., Mantovani, R.G., Mastelini, S.M., Carvalho, A.C. de, Campos, G.F. and Junior, S.B. 2019. A meta-learning approach for selecting image segmentation algorithm. *Pattern Recognition Letters*. 128, (2019), 480–487.
[2] Aha, D.W., Kibler, D. and Albert, M.K. 1991. Instance-based learning algorithms. *Machine learning*. 6, 1 (1991), 37–66.
[3] Beel, J. and Brunel, V. 2019. Data Pruning in Recommender Systems Research: Best-Practice or Malpractice. *ACM RecSys*. (2019).
[4] Beel, J., Gipp, B., Langer, S. and Breitinger, C. 2016. Research-paper recommender systems: a literature survey. *International Journal on Digital Libraries*. 17, 4 (Nov. 2016), 305–338.
[5] Beel, J., Langer, S., Nürnberger, A. and Genzmehr, M. 2013. The Impact of Demographics (Age and Gender) and Other User-Characteristics on Evaluating Recommender Systems. *Research and Advanced Technology for Digital Libraries - International Conference on Theory and Practice of Digital Libraries, TPDL 2013, Valletta, Malta, September 22-26, 2013. Proceedings* (2013), 396–400.
[6] Beel, J., Smyth, B. and Collins, A. 2019. RARD II: The 94 Million Related-Article Recommendation Dataset. *Proceedings of the 1st Interdisciplinary Workshop on Algorithm Selection and Meta-Learning in Information Retrieval (AMIR)* (2019).

[7] Beel, J., Tyrell, B., Bergman, E., Collins, A. and Nagoor, S. 2020. Siamese Meta-Learning and Algorithm Selection with'Algorithm-Performance Personas'[Proposal]. *arXiv preprint arXiv:2006.12328*. (2020).

[8] Bontrager, P., Khalifa, A., Mendes, A. and Togelius, J. 2016. Matching games and algorithms for general video game playing. *Twelfth Artificial Intelligence and Interactive Digital Entertainment Conference* (2016).

[9] Collins, A. and Beel, J. 2019. A First Analysis of Meta-Learned Per-Instance Algorithm Selection in Scholarly Recommender Systems. *3rd Workshop on Recommendation in Complex Scenarios (ComplexRec 2019), in conjunction with the 13th ACM Conference on Recommender Systems (RecSys 2019)* (2019).

[10] Collins, A. and Beel, J. 2019. Document Embeddings vs. Keyphrases vs. Terms for Recommender Systems: A Large-Scale Online Evaluation. *2019 ACM/IEEE Joint Conference on Digital Libraries (JCDL)* (2019), 130–133.

[11] Collins, A., Tkaczyk, D. and Beel, J. 2018. A Novel Approach to Recommendation Algorithm Selection using Meta-Learning. *26th AIAI Irish Conference on Artificial Intelligence and Cognitive Science*. (2018), 210–219.

[12] Cunha, T., Soares, C. and Carvalho, A.C. de 2018. CF4CF: recommending collaborative filtering algorithms using collaborative filtering. *Proceedings of the 12th ACM Conference on Recommender Systems* (2018), 357–361.

[13] Cunha, T., Soares, C. and Carvalho, A.C. de 2018. Metalearning and Recommender Systems: A literature review and empirical study on the algorithm selection problem for Collaborative Filtering. *Information Sciences*. 423, (2018), 128–144.

[14] Cunha, T.D.S. 2019. Recommending Recommender Systems: tackling the Collaborative Filtering algorithm selection problem. (2019).

[15] Edenhofer, G., Collins, A., Aizawa, A. and Beel, J. 2019. Augmenting the DonorsChoose.org Corpus for Meta-Learning. *1st Interdisciplinary Workshop on Algorithm Selection and Meta-Learning in Information Retrieval (AMIR)*. (2019).

[16] Ekstrand, M. and Riedl, J. 2012. When recommenders fail: predicting recommender failure for algorithm selection and combination. *Proceedings of the sixth ACM conference on Recommender systems* (2012), 233–236.

[17] Ekstrand, M.D. and Pera, M.S. 2017. The Demographics of Cool: Popularity and Recommender Performance for Different Groups of Users. *Proceedings of the Poster Track of the 11th ACM Conference on Recommender Systems (RecSys 2017), Como, Italy, August 28, 2017.* (2017).

[18] Ferrari, D.G. and Castro, L.N. de 2012. Clustering algorithm recommendation: a meta-learning approach. *International Conference on Swarm, Evolutionary, and Memetic Computing* (2012), 143–150.

[19] Hernández-Lobato, D., Martı́Nez-MuñOz, G. and Suárez, A. 2013. How large should ensembles of classifiers be? *Pattern Recognition*. 46, 5 (2013), 1323–1336.

[20] Houstis, E.N., Catlin, A.C., Dhanjani, N., Rice, J.R., Ramakrishnan, N. and Verykios, V. 2002. MyPYTHIA: a recommendation portal for scientific software and services. *Concurrency and Computation: Practice and Experience*. 14, 13-15 (2002), 1481–1505.

[21] Hug, N. 2017. Surprise, a Python library for recommender systems.

[22] Ke, G., Meng, Q., Finley, T., Wang, T., Chen, W., Ma, W., Ye, Q. and Liu, T.-Y. 2017. Lightgbm: A highly efficient gradient boosting decision tree. *Advances in neural information processing systems* (2017), 3146–3154.

[23] Kerschke, P., Hoos, H.H., Neumann, F. and Trautmann, H. 2019. Automated algorithm selection: Survey and perspectives. *Evolutionary computation*. 27, 1 (2019), 3–45.

[24] Kotthoff, L. 2016. Algorithm selection for combinatorial search problems: A survey. *Data Mining and Constraint Programming*. Springer. 149–190.

[25] Kuncheva, L.I. 2000. Clustering-and-selection model for classifier combination. *KES'2000. Fourth International Conference on Knowledge-Based Intelligent Engineering Systems and Allied Technologies. Proceedings (Cat. No. 00TH8516)* (2000), 185–188.

[26] Lee, J.W. and Giraud-Carrier, C. 2013. Automatic selection of classification learning algorithms for data mining practitioners. *Intelligent Data Analysis*. 17, 4 (2013), 665–678.

[27] Ler, D., Teng, H., He, Y. and Gidijala, R. 2018. Algorithm Selection for Classification Problems via Cluster-based Meta-features. *2018 IEEE International Conference on Big Data (Big Data)* (2018), 4952–4960.

[28] Nechaev, A., Zhukova, N. and Meltsov, V. 2019. Utilizing Metadata to Select a Recommendation Algorithm for a User or an Item. *11th Majorov International Conference on Software Engineering and Computer Systems (MICSECS 2019)*. (2019).

[29] Opitz, D. and Maclin, R. 1999. Popular ensemble methods: An empirical study. *Journal of artificial intelligence research*. 11, (1999), 169–198.

[30] Pedregosa, F. et al. 2011. Scikit-learn: Machine Learning in Python. *Journal of Machine Learning Research*. 12, (2011), 2825–2830.

[31] Pimentel, B.A. and Carvalho, A.C. de 2019. A new data characterization for selecting clustering algorithms using meta-learning. *Information Sciences*. 477, (2019), 203–219.

[32] Rendell, L., Seshu, R. and Tcheng, D. 1987. More Robust Concept Learning Using Dynamically-Variable Bias. *Proceedings of the Fourth International Workshop on Machine Learning* (1987), 66–78.

[33] Rivolli, A., Garcia, L.P., Soares, C., Vanschoren, J. and Carvalho, A.C. de 2018. Characterizing classification datasets: a study of meta-features for meta-learning. *arXiv preprint arXiv:1808.10406*. (2018).

[34] Smith-Miles, K.A. 2009. Cross-disciplinary perspectives on meta-learning for algorithm selection. *ACM Computing Surveys (CSUR)*. 41, 1 (2009), 6.

[35] De Souto, M.C., Prudencio, R.B., Soares, R.G., De Araujo, D.S., Costa, I.G., Ludermir, T.B. and Schliep, A. 2008. Ranking and selecting clustering algorithms using a meta-learning approach. *2008 IEEE International Joint Conference on Neural Networks (IEEE World Congress on Computational Intelligence)* (2008), 3729–3735.

[36] Tyrrell, B., Bergman, E., Jones, G.J. and Beel, J. 2020. Algorithm-Performance Personas' for Siamese Meta-Learning and Automated Algorithm Selection. *7th ICML Workshop on Automated Machine Learning (AutoML 2020)*. (2020).